Projecting the optimal control strategy on invasive plants combining effects of herbivores and native plants resistance

Zhiyuan Fu[1], Yuanming Lu[2], Donald DeAngelis[2], and Bo Zhang[3,4*]

[1]Co-Innovation Center for Sustainable Forestry in Southern China, Jiangsu Province Key Laboratory of Soil and Water Conservation and Ecological Restoration, Nanjing Forestry University, Nanjing, China

[2]Department of Biology, University of Miami, Coral Gables, Florida, USA

[3]Department of Environmental Science and Policy, University of California, Davis, California, USA

[4]Department of Natural Ecology Resource and Management, Oklahoma State University, Stillwater, Oklahoma, USA

*Corresponding author: bozhangophelia@gmial.com (Bo Zhang)

**Abstract**

Understanding how to limit biological invasion is critical, especially in the context of accelerating anthropogenic ecological changes. Although biological invasion success could be explained by the lack of natural enemies in new regions, recent studies have revealed that resident herbivores often do have a substantial effect on both native and invasive plants. Very few studies have included consideration of native plant resistance while estimating methods of controlling invasion; hence, it is unclear to what extent the interactive effects of controlling approaches and native plants' resistance could slow down or even inhibit biological invasion. We developed a spatial modeling framework, using a paired logistic equation model, with considerations of the dispersal processes, to capture the dynamics change of native and invasive plants under various strategies of control. We found that when biocontrol agents could have a strong effect on invasive plant, that could almost completely limit the invasion, together with a high native plant resistance. However, a high application frequency is needed make an efficient impact, whereas, a low frequency treatment leads to nearly the same outcome as the no treatment case. Lastly, we showed that evenly controlling a larger area with a weaker effect still lead to a better outcome than focusing on small patches with a stronger effect. Overall, this study has some management implications, such as how to determine the optimal allocation strategy.

**Introduction**

The past few centuries have witnessed a spike in the number of alien plants invading and establishing self-sustaining populations in new regions; often driving local biodiversity loss and habitat degradation (Mainka and Howard 2010, Simberloff et al. 2013, Zhang et al. 2017, Zhang et al. 2018a). One possible explanation of biological invasion success is the lack of natural enemies of the plants in new regions (Keane and Crawley 2002). However, several studies, based on a broad plant invasions studies, have revealed that resident herbivores often do have a substantial effect in reducing the establishment and performance of colonizing invaders as well (Levine et al. 2004, Zhang et al. 2018b). Specialist pathogens or herbivores (e.g. biocontrol agents), are often introduced from the home range of invasive weeds to reduce the dominance of invaders and hence facilitate the recovery of native plants (Zhang et al. 2017), but it is possible that more use could be made of the resident herbivores through appropriately supplementing resident herbivore density within pockets of weedy invaders. Indeed, it would be pretty costly to raise large enough amounts of native herbivores to wipe out invading and native plants, hence, substituting herbicide for native herbivores could be another option.

Although the competitive ability of common native plants might be not be sufficient to resist the invasive plant, it could also be a factor that should be considered in efforts to control alien invasion through supplementing native herbivores (Zhang and van Kleunen 2019). Together with the interaction of controlling approaches (e.g. resident herbivores (target on both native and invasive plants) or herbicide (target on both native and invasive plants) or specialist biocontrol agents (target only on invasive plants)), native plants' biotic resistance may have the potential to inhibit, or even alter invasion outcomes. For instance, a recent field study has found that some native plants still co-occupy similar niches as invasive ones, especially in late-successional communities (Golivets and Wallin 2018). Nonetheless, very few studies have included consideration of native plant resistance while estimating methods of controlling invasion (Kettenring and Adams 2011, Epanchin-Niell and Wilen 2012). Furthermore, as controlling approaches' impacts are heterogeneous in spatiotemporal gradients, and because the relative competitive abilities of the invasive and native plants could vary along environmental gradients, it is unclear to what extent the interactive effects of controlling approaches and native plants resistance could slow down or even inhibit biological invasion.

More importantly, understanding the possibility and efficiency of utilizing controlling approaches, in addition to native plants' biological resistance, could provide important field management suggestions. For example, such understanding will assist a management dilemma of prioritizing of how best to deploy limited resources among different regions and what is the best application frequency of controlling approaches when invasive populations occur simultaneously in different but interconnected regions (Ferguson et al. 2001, Keeling et al. 2001, Dye and Gay 2003). Therefore, developing a modeling framework that captures interactive effects of different controlling approaches and native plants' biological resistance on limiting invasion spread is essential. More importantly, most previous models lacked inclusion of the dispersal process, so that it is unclear how dispersal rates of both native and invasive plants, and environmental spatial heterogeneity alter the outcome of various control strategies (Hastings et al. 2006, Blackwood et al. 2010, Epanchin-Niell and Hastings 2010, Bonneau et al. 2019).

To develop a spatial modeling framework that can capture the dynamics of native and invasive plants, we used a paired logistic equation model with considerations of the dispersal processes and competition between native and invasive plants. Specifically, both populations can disperse from a given cell to the neighboring four cells. We further simulated the effects of local herbivores or herbicide in a same way of using a same mortality rate on both native and invasive plants, while the natural enemy (e.g. biocontrol agents) solely kills invasive plants at a different mortality rate. We used this model to test three hypotheses: 1. Together with native plants' resistance, applying a natural enemy (e.g. biocontrol agents) is a more efficient management way than using the supplementing of local herbivores or herbicide on controlling invasive plants; 2. Controlling on a smaller region each time but more frequently is better than controlling in a bigger area but in a lower frequency; 3. The optimal control strategy varies with plants' intrinsic growth rate, competition coefficient and dispersal rate.

**Methods and materials**

**Model framework** - We propose a paired logistic equation model describing dispersing native ($N$) and invasive ($I$) populations in a $100 \times 100$ spatial environment:

$$\frac{dN_{i,j}}{dt} = rN_{i,j}\left(\frac{K_{i,j}-N_{i,j}-\alpha_{21}I_{i,j}}{K_{i,j}}\right) - dN_{i,j} + \frac{1}{4}dN_{i,(j+1)} + \frac{1}{4}dN_{i,(j-1)} + \frac{1}{4}dN_{(i+1),j} + \frac{1}{4}dN_{(i-1),j} \quad (1a)$$

$$\frac{dI_{i,j}}{dt} = RI_{i,j}\left(\frac{K_{i,j}-I_{i,j}-\alpha_{12}N_{i,j}}{K_{i,j}}\right) - DI_{i,j} + \frac{1}{4}DI_{i,(j+1)} + \frac{1}{4}DI_{i,(j-1)} + \frac{1}{4}DI_{(i+1),j} +$$
$$\frac{1}{4}DI_{(i-1),j} \qquad (1b)$$

where *i* and *j* are the row and column numbers of the cell, *r (R)* is intrinsic growth rate of native (invasive) population, $K_{i,j}$ is the carrying capacity of cell(*i,j*), $\alpha_{21}$ is the competition coefficient of invasive on native, $\alpha_{12}$ is the competition coefficient of native on invasive. and *d(D)* is the dispersal rate of native (invasive) population.

**Initial conditions** - Native populations are assumed to initially occupy all cells and each cell is assumed to have a randomly determined initial population size ($N_0$ between 0 and 1). An initial invasive population is assumed to occupy only the center cell ($I_0$ = 2), see Fig.1 for more details.

**Invasion process** – Every time step, one new invasive population was added in a randomly selected cell. This process represents the assumption that invasive species has a chance to randomly disperse to a new region on each time step.

**Dispersal process** – Every time step, a proportion of *d(D)* native (invasive) populations dispersed from each cell, then allocated equally among the four neighboring cells (up, down, left, and right). Note that we set the outer two circles as buffer zone, so that they received dispersed populations but did not disperse back. We did not include populations in buffer zone when calculated total population abundance.

**Bio-control treatment** – Bio-control treatment was simulated as a mortality rate only on invasive populations in the target cell.

**Local herbivore or herbicide control** –Control treatment by local herbivores or herbicide in each cell was simulated as a process that killed both native and invasive populations.

**Controlling pattern** – Two patterns for both bio-control and local herbivore or herbicide control groups were performed: 1. A random cell with a non-zero invasive population was selected at every time step, and the whole row where the selected cell was located would be treated. For instance, if it is the bio-control case, then all invasive populations within that row would be treated, if it is the local herbivore or herbicide case, all native and invasive populations in that row would be treated. 2. A random cell with a non-zero invasive population was selected at every time step, and a 5 × 5 cells in which the selected cell was in the center, would be treated.

**Parameter estimation** - The system characteristics represented by parameter values in the model were estimated based on previous findings of general patterns between invasive and native species. For instance, invasive species generally have higher competition coefficients than natives (Vila and Weiner 2004), and invasives have overall stronger dispersal ability than natives (Nunez‐Mir et al. 2019) (see more information listed in Table 1). Indeed, invasive plants often have higher growth rate than natives, but we used the same growth rate for both species because the main focus of this study is to look at the role of competition coefficient and controlling treatments. Additionally, a higher growth rate of invasive species is believed to further favor the invasion process, so only has an additional quantitative effect.

**Scenarios**

**Scenario 1: Comparison of biocontrol and local herbivore or herbicide treatments with different levels of native plant biotic resistance.** In every time step,

a cell with a non-zero invasive population was randomly selected. Bio-control treatment was simulated as either completely eliminating all invasive populations in the whole row where the selected cell was located or removing or invasive populations in a 5 × 5 cells in which the selected cell was in the center. Local herbivory or herbicide was simulated as completely eliminating all populations of both invasive and native plants. In both treatments, we performed two levels of native plant biotic resistance, low ($\alpha_{12}$ = 1) and high ($\alpha_{12}$ = 1.1). In each sub-scenario, we simulated three treatment frequencies, which were no treatment, high treatment (controlling treatment was applied in every step), and low frequency (controlling treatment was applied in every 100 steps). Each sub-scenario ran for 50 times because there are several stochastic processes involved in this model; e.g., one invasive population was added to a randomly selected cell in each time step. We calculated the total proportion of (Invasive) = $\frac{sum(I_{(i,j)})}{sum(N_{(i,j)})+sum(I_{(i,j)})}$ to represent the landscape dominance of the invasive species.

**Scenario 2: Comparison of biocontrol and local herbivore or herbicide treatments with different controlling area sizes and mortality rates.** Under both bio-control and local herbivore or herbicide treatments as described in the previous scenario, we conducted two cases: 1. size of different areas controlled (big: eliminated populations in the whole row where the selected cell was located; small: eliminated populations in 1/3 of the cells within that row); and 2. different mortality rates (high: killed all target populations; low: killed half of target populations). Similarly, three treatment frequencies were simulated in each sub-scenario, which were no treatment, high treatment, and low frequency. Each sub-scenario ran for 50 times.

**Scenario 3: Roles of intrinsic growth rate and competition coefficient on altering controlling efficiency.** To capture a more complete information on the controlling efficiency of bio-control and the local herbivore, we simulated the change of final total proportion of invasives under a series of combinations of intrinsic growth rate ($r = R =$ [0.1:0.1:0.5]), and competition coefficient of natives on invasives ($\alpha_{12}$ = 0.8, 1, 1.2, 1.4) under each treatment. Similarly, three treatment frequencies were simulated in each sub-scenario, which were no treatment, high treatment frequency, and low frequency. We ran each combination for 50 times and used the average value to represent each condition.

**Test of more parameter values** – Different values were applied to two parameters regarding to the controlling efficiency under both biocontrol and local herbivore treatments as described in the previous scenarios: 1. Initial proportions of native population to carrying capacity (0.1, 0.2, 0.5 and 0.9). In this case, each cell had a same initial native population based on the proportion; 2. Dispersal rates of invasive species ($D$) (0.1, 0.2, 0.5 and 0.8). We are aware that 0.5 and 0.8 might be a much higher dispersal rate than realistic, but the main objective here is to explore the impact of a broad range of dispersal rate. Treatment frequency was set as every five-time steps. Each condition ran for 50 times.

**Results**

We compared the efficiency control of using bio-control and local herbivores or herbicide, as completely eliminating all populations in the whole row where the selected cell was located. Overall, under the same conditions, treatments with high native plant resistance (Fig. 2 d-f, and j-l) resulted in lower proportions of invasive population for the biocontrol than for local herbivores, whereas, there were no obvious difference between the cases of no treatments (Fig. 2 a, d, g, j) and low frequency treatments (Fig. 2 c, f, i, l) when other conditions were same. We noticed that treatment with a high frequency led to lower proportions of invasive population in all the sub-scenarios. In particular, using biocontrol in a high frequency when native species also had high resistance can almost completely wipe out invasive populations (Fig. 2e). Even with low native plant resistance, applying bio-control in a high frequency still led to a relative low proportion of the invasive population (< 0.4) (Fig. 2b). Compared with the high efficiency of biocontrol, the local herbivores or herbicide treatment with low native resistance was incapable of limiting invasion, possibly because this type of treatment attacked equally on both invasive and native populations, and invasive populations with higher competition ability could take advantage of the herbivores' effect on native populations (Fig. 2h). When the corresponding treatment was applied on the populations in a 5 × 5 cells in which the selected cell was in the center, we found similar trends as another controlling pattern (Fig. S1 in the Appendix).

In Scenario 2, we compared the effects of both the different sizes to which control was applied and different mortality rates due to control. Similar to the previous scenario, we found that there was no obvious difference between the effects of biocontrol and that of local herbivores or herbicide under the cases of no treatments (Fig. 3 a, d, g, j) and low frequency treatments (Fig. 2 c, f, i, l) when other conditions were the same, regardless of area to which control was applied and the efficiency. With biocontrol treatment, controlling over a larger area with low mortality led to a much lower proportion of invasive population (< 0.2) (Fig. 3b) than focusing on a small area, even though a higher mortality rate was assumed for the smaller area (Fig. 3e); whereas, with local herbivores treatment, we did not see obvious difference between different treating area and mortality rates (Fig. 3h and k).

In Scenario 3 we still kept the same intrinsic growth rate for both invasive and native species, but the rate was varied from 0.1 to 0.5. Additionally, competition coefficients of native on invasive ($\alpha_{12}$) varied from 0.8 to 1.4, which were still lower than competition coefficients of the invasive species ($\alpha_{21} = 2$). In general, we found that the final proportion of invasive population changed more sensitively to the change of $\alpha_{12}$ than to $r$. When $\alpha_{12}$ was low, invasive species could completely dominate in all treatments (dark blue boxes in Fig. 4). When $\alpha_{12}$ increased, bio-control was able to limit invasion when $r$ was small and treatment frequency was high (blue boxes in Fig. 4b). On the contrary, local herbivores or herbicide could not limit invasion under the same condition (blue boxes in Fig. 4e). When $\alpha_{12}$ was increased further, while remaining lower than $\alpha_{21}$, both biocontrol and local herbivores or herbicide could limit invasion in most cases (green and yellow boxes in Fig. 4). It is interesting to note that with a high value of $\alpha_{12}$, native species could outcompete invasive ones even without the control treatment (green and yellow boxes in Fig. 4a and d).

Compared to the impact of changes of $\alpha_{12}$ and $r$, we found that the change of the resulting proportion of invasive population was less sensitive to the initial proportion of native species and dispersal rate (Fig. 5). Together, these results indicate that the intrinsic growth rate and competition coefficients of native species were the most influential factors on determining invasion success.

**Discussion**

Biocontrol agents have made important contributions on controlling plant invasion (Fravel 2005). Due to the fact that these agents only target the specific invasive plants, this study showed that successful biocontrol agents when having a strong effect on invasive plant could almost completely limit the invasion, together with a high native plant resistance. This results, to some degree, support the enemy release hypothesis – the success of invasive plants is due to escaping from their natural enemies (Keane and Crawley 2002). Hence, theoretically, the decline in enemy release strength can lead to the decrease in invasive abundance (Keane and Crawley 2002), whereas, as some empirical data have suggested, reduction in herbivory is another key mechanism for invasive plants to be successful (Engelkes et al. 2012).

Application frequency of control also plays an important role on determining the control efficiency, such that only a high application frequency could make an efficient impact, whereas, a low frequency treatment leads to nearly the same outcome as the no treatment case. This result is consistent to some field studies that found a high frequency treatment (e.g. clipping, fire) generally led to the best controlling outcome (Emery and Gross 2005, Tang et al. 2009, Tang et al. 2010, Valentine et al. 2012).

Additionally, this result has some management implications, such as how to determine the optimal allocation strategy. For instance, this study refers back an important management question on determining which strategy of control is better: control a larger area with a treatment that leads to a low mortality rate or focus on smaller areas but using a stronger treatment (high mortality rate)? We showed that with biocontrol, applying control to a larger area, yet with a low mortality rate, is still more efficient than vice versa. Consistent with previous results, biocontrol did a better job than local herbivores in this scenario. This result also agrees with previous simulation and empirical studies that showed that evenly controlling a larger area is better than focusing on small patches (Moody and Mack 1988, Arroyo-Esquivel et al. 2019, Zhang et al. 2020). Our results, which suggest there is no significant difference between treating area and mortality rate with local herbivores treatment, lead to the conclusion that resident herbivores, which prefer native plants, have little impact on invasion process (Joshi and Vrieling 2005). Therefore, the key factor of the effective bio-control is to invader-specific herbivores with frequent treatments (Van Dyken and Zhang 2018, Van Dyken and Zhang 2019, Zhang et al. 2020, Zhang and DeAngelis 2020).

Understanding how the outcome of control changes according to the change of vegetation characteristics is essential (Hobbs and Humphries 1995). Here, our results reveal that with an increase of native plants' competitive ability, treatment such as biocontrol has a great possibility of success in limiting invasion. This result suggests the importance of maintaining the health of native vegetation and high biodiversity to slow down invasion (Early et al. 2016, Pile et al. 2019). Notably, community diversity and evenness could help to reduce invasion success. The healthy vegetation with high biodiversity level has larger resistance to disturbance, which in turn decreases invasibility (Hooper et al. 2005). In essence, the invasion process restructures the richness of seedlings and saplings in the community, resulting in increasing invasibility of abandoned patches but less so in secondary forest (Mullah et al. 2014). Our theoretical results show the overall lower effectiveness of biocontrol of an invaded community with low resistance compared to that with high resistance. The effectiveness of bio-control treatment therefore relies on a healthy vegetation community. Furthermore, we did not find a significant difference between having different initial density of native plants or various dispersal rates.

**Acknowledgement**

This work was supported by UC Davis Chancellors' postdoc fellowship to BZ, Greater Everglades Priority Ecosystem Science program to YL and DLD. Any use of trade, firm, or product names is for descriptive purposes only and does not imply endorsement by the U.S. Government.

**Declarations**

**Funding**: This work was supported by UC Davis Chancellors' postdoc fellowship to BZ, Greater Everglades Priority Ecosystem Science program to YL and DLD.

**Conflicts of interest/Competing interests**: Not applicable

**Availability of data and material**: Not applicable

**Code availability**: Will be uploaded once accepted

**Authors' contributions:** ZF and BZ designed the study, YL and BZ performed the simulations, all authors wrote the paper.

## References

Arroyo-Esquivel, J., F. Sanchez, and L. A. Barboza. 2019. Infection model for analyzing biological control of coffee rust using bacterial anti-fungal compounds. Mathematical Biosciences **307**:13-24.

Blackwood, J., A. Hastings, and C. Costello. 2010. Cost-effective management of invasive species using linear-quadratic control. Ecological Economics **69**:519-527.

Bonneau, M., J. Martin, N. Peyrard, L. Rodgers, C. M. Rornagosa, and F. A. Johnson. 2019. Optimal spatial allocation of control effort to manage invasives in the face of imperfect detection and misclassification. Ecological Modelling **392**:108-116.

Dye, C., and N. Gay. 2003. Modeling the SARS epidemic. Science **300**:1884-1885.

Early, R., B. A. Bradley, J. S. Dukes, J. J. Lawler, J. D. Olden, D. M. Blumenthal, P. Gonzalez, E. D. Grosholz, I. Ibañez, and L. P. Miller. 2016. Global threats from invasive alien species in the twenty-first century and national response capacities. Nature Communications **7**:1-9.

Emery, S. M., and K. L. Gross. 2005. Effects of timing of prescribed fire on the demography of an invasive plant, spotted knapweed Centaurea maculosa. Journal of Applied Ecology **42**:60-69.

Engelkes, T., B. Wouters, T. M. Bezemer, J. A. Harvey, and W. H. van der Putten. 2012. Contrasting patterns of herbivore and predator pressure on invasive and native plants. Basic and Applied Ecology **13**:725-734.

Epanchin-Niell, R. S., and A. Hastings. 2010. Controlling established invaders: integrating economics and spread dynamics to determine optimal management. Ecology Letters **13**:528-541.

Epanchin-Niell, R. S., and J. E. Wilen. 2012. Optimal spatial control of biological invasions. Journal of Environmental Economics and Management **63**:260-270.

Ferguson, N. M., C. A. Donnelly, and R. M. Anderson. 2001. The foot-and-mouth epidemic in Great Britain: Pattern of spread and impact of interventions. Science **292**:1155-1160.

Fravel, D. 2005. Commercialization and implementation of biocontrol. Annu. Rev. Phytopathol. **43**:337-359.

Golivets, M., and K. F. Wallin. 2018. Neighbour tolerance, not suppression, provides competitive advantage to non-native plants. Ecology Letters **21**:745-759.

Hastings, A., R. J. Hall, and C. M. Taylor. 2006. A simple approach to optimal control of invasive species. Theoretical Population Biology **70**:431-435.

Hobbs, R. J., and S. E. Humphries. 1995. An integrated approach to the ecology and management of plant invasions. Conservation Biology **9**:761-770.

Hooper, D. U., F. Chapin Iii, J. Ewel, A. Hector, P. Inchausti, S. Lavorel, J. H. Lawton, D. Lodge, M. Loreau, and S. Naeem. 2005. Effects of biodiversity on ecosystem functioning: a consensus of current knowledge. Ecological monographs **75**:3-35.

Joshi, J., and K. Vrieling. 2005. The enemy release and EICA hypothesis revisited: incorporating the fundamental difference between specialist and generalist herbivores. Ecology Letters **8**:704-714.

Keane, R. M., and M. J. Crawley. 2002. Exotic plant invasions and the enemy release hypothesis. Trends in Ecology & Evolution **17**:164-170.

Keeling, M. J., M. E. J. Woolhouse, D. J. Shaw, L. Matthews, M. Chase-Topping, D. T. Haydon, S. J. Cornell, J. Kappey, J. Wilesmith, and B. T. Grenfell. 2001. Dynamics of the 2001 UK foot and mouth epidemic: Stochastic dispersal in a heterogeneous landscape. Science **294**:813-817.

Kettenring, K. M., and C. R. Adams. 2011. Lessons learned from invasive plant control experiments: a systematic review and meta-analysis. Journal of Applied Ecology **48**:970-979.

Levine, J. M., P. B. Adler, and S. G. Yelenik. 2004. A meta-analysis of biotic resistance to exotic plant invasions. Ecology Letters **7**:975-989.

Mainka, S. A., and G. W. Howard. 2010. Climate change and invasive species: double jeopardy. Integrative Zoology **5**:102-111.

Moody, M. E., and R. N. Mack. 1988. Controlling the spread of plant invasions: the importance of nascent foci. Journal of Applied Ecology:1009-1021.

Mullah, C. J. A., K. Klanderud, Ø. Totland, and D. Odee. 2014. Community invasibility and invasion by non-native Fraxinus pennsylvanica trees in a degraded tropical forest. Biological invasions **16**:2747-2755.

Nunez-Mir, G. C., Q. Guo, M. Rejmánek, B. V. Iannone III, and S. Fei. 2019. Predicting invasiveness of exotic woody species using a traits-based framework. Ecology:e02797.

Pile, L. S., L. Vickers, M. Stambaugh, C. Norman, and G. G. Wang. 2019. The tortoise and the hare: A race between native tree species and the invasive Chinese tallow. Forest Ecology and Management **445**:110-121.

Simberloff, D., J.-L. Martin, P. Genovesi, V. Maris, D. A. Wardle, J. Aronson, F. Courchamp, B. Galil, E. García-Berthou, and M. Pascal. 2013. Impacts of biological invasions: what's what and the way forward. Trends in Ecology & Evolution **28**:58-66.

Tang, L., Y. Gao, C. Wang, J. Wang, B. Li, J. Chen, and B. Zhao. 2010. How tidal regime and treatment timing influence the clipping frequency for controlling invasive Spartina alterniflora: implications for reducing management costs. Biological invasions **12**:593-601.

Tang, L., Y. Gao, J. Wang, C. Wang, B. Li, J. Chen, and B. Zhao. 2009. Designing an effective clipping regime for controlling the invasive plant Spartina alterniflora in an estuarine salt marsh. Ecological Engineering **35**:874-881.

Valentine, L. E., L. Schwarzkopf, and C. N. Johnson. 2012. Effects of a short fire-return interval on resources and assemblage structure of birds in a tropical savanna. Austral Ecology **37**:23-34.

Van Dyken, J. D., and B. Zhang. 2018. Carrying capacity of a spatially-structured population: disentangling the effects of dispersal, growth parameters, habitat heterogeneity and habitat clustering. Journal of Theoretical Biology.

Van Dyken, J. D., and B. Zhang. 2019. Carrying capacity of a spatially-structured population: Disentangling the effects of dispersal, growth parameters, habitat heterogeneity and habitat clustering. Journal of Theoretical Biology **460**:115-124.

Vila, M., and J. Weiner. 2004. Are invasive plant species better competitors than native plant

species?–evidence from pair-wise experiments. Oikos **105**:229-238.

Zhang, B., D. DeAngelis, W.-M. Ni, Y. Wang, L. Zhai, A. Kula, S. Xu, and D. V. Dyken. 2020. Effect of stressors on the carrying capacity of spatially distributed metapopulations. American Naturalist.

Zhang, B., and D. L. DeAngelis. 2020. An overview of agent-based models in plant biology and ecology. Annals of Botany.

Zhang, B., D. L. DeAngelis, M. B. Rayamajhi, and D. Botkin. 2017. Modeling the long-term effects of introduced herbivores on the spread of an invasive tree. Landscape Ecology **32**:1147-1161.

Zhang, B., X. Liu, D. L. DeAngelis, L. Zhai, M. B. Rayamajhi, and S. Ju. 2018a. Modeling the compensatory response of an invasive tree to specialist insect herbivory. Biological Control **117**:128-136.

Zhang, Z., X. Pan, D. Blumenthal, M. van Kleunen, M. Liu, and B. Li. 2018b. Contrasting effects of specialist and generalist herbivores on resistance evolution in invasive plants. Ecology **99**:866-875.

Zhang, Z., and M. van Kleunen. 2019. Common alien plants are more competitive than rare natives but not than common natives. Ecology Letters **22**:1378-1386.

**Figures**

Figure 1. (a) Schematic figure of the model. a. resident herbivores were simulated as a negative effect on both native and invasive plants while bio-control agents only targeted invasive plants. (b) we developed a 100 × 100 spatial environment, in which the outer two circles were set as buffer zone, so that they received dispersed populations but did not disperse back. Native plants started in all cells in the beginning, while invasive only occurred in the center cell. Every time step, a proportion of *d(D)* native (invasive) populations dispersed from each cell, then allocated equally among the four neighboring cells (up, down, left, and right).

Figure 2. Proportions of invasive population changes over time with bio-control treatment (a-f) and local herbivores (g-l), low resistance $\alpha_{12} = 1$, high resistance $\alpha_{12} = 1.1$. r = R = 0.2, $\alpha_{21} = 2$, d = 0.1 and D = 0.2. Black lines: no treatment applied; red lines: treatment applied in every one-time step; green lines: treatment applied in every 100-time steps. This is the case when treatment was simulated as completely eliminating all populations in the whole row where the selected cell was located.

Figure 3. Proportions of invasive population changes over time with bio-control treatment in small area but high mortality (a-c), bio-control treatment in big area but low mortality (d-f), local herbivores in small area but high mortality (g-i), and local herbivores in big area but low mortality (j-l). Black lines: no treatment applied; red lines: treatment applied in every one-time step; green lines: treatment applied in every 100-time steps.

Figure 4. Final proportions of invasive population changes over time with bio-control treatment (a-c) and local herbivores (d-f). $r = R = [0.1:0.1:0.5]$, $\alpha_{12} = [0.8, 1, 1.2, 1.4]$, $\alpha_{21} = 2, d = 0.1, D = 0.2$. Left panels: no treatment applied; middle panels: treatment applied in every one-time step; right panels: treatment applied in every 100-time steps.

Figure 5. Changes in the proportion of invasive population over time, when initial proportion of native population to carrying capacity changed between 0.1, 0.2, 0.5 and 0.9 (a-d, i-l); dispersal rate of invasive species (*D*) changed between 0.1, 0.2, 0.5 and 0.8 (e-h, m-p). Top two rows represented bio-control treatment and bottom two rows represented local herbivores treatments.

Figure 1

a 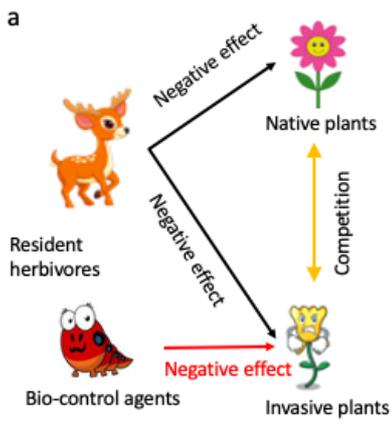

b 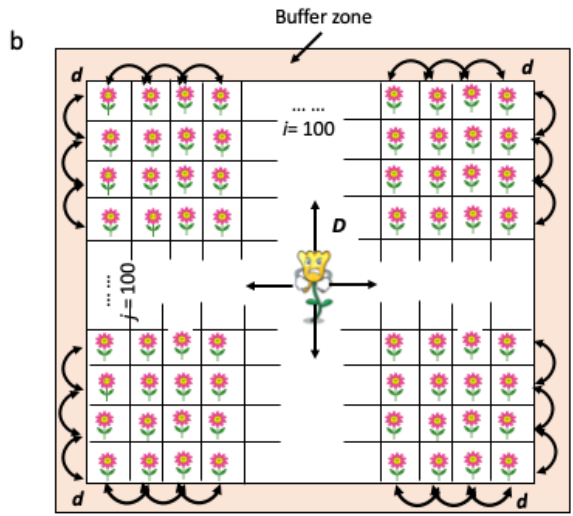

$$\frac{dN}{dt} = r * N * \frac{(K - N - \alpha 21 * I)}{K} + f(d)$$

$$\frac{dI}{dt} = R * I * \frac{(K - I - \alpha 12 * N)}{K} + f(D)$$

Figure 2

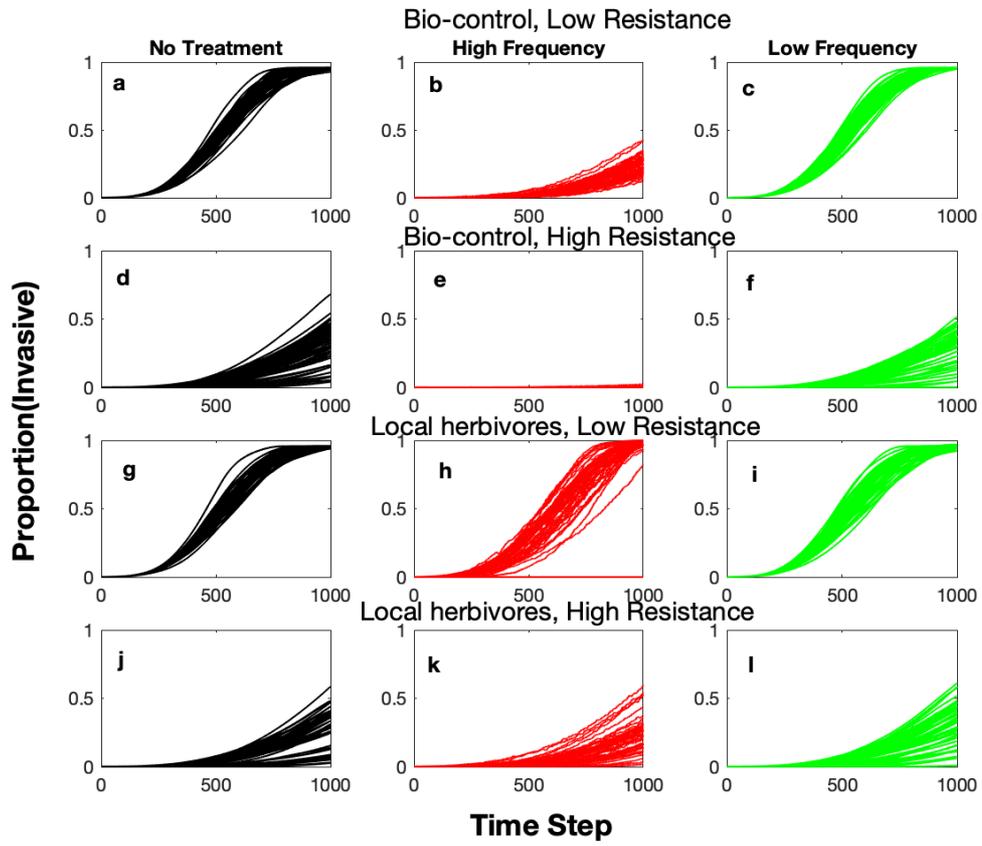

Figure 3

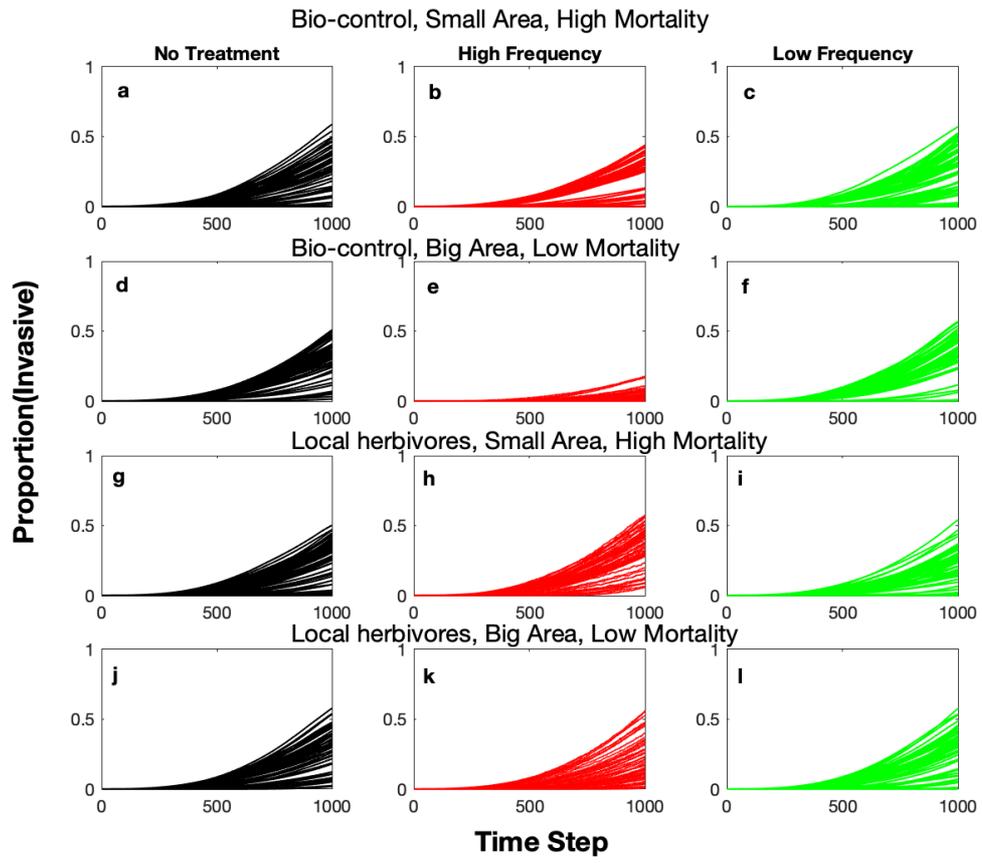

Figure 4

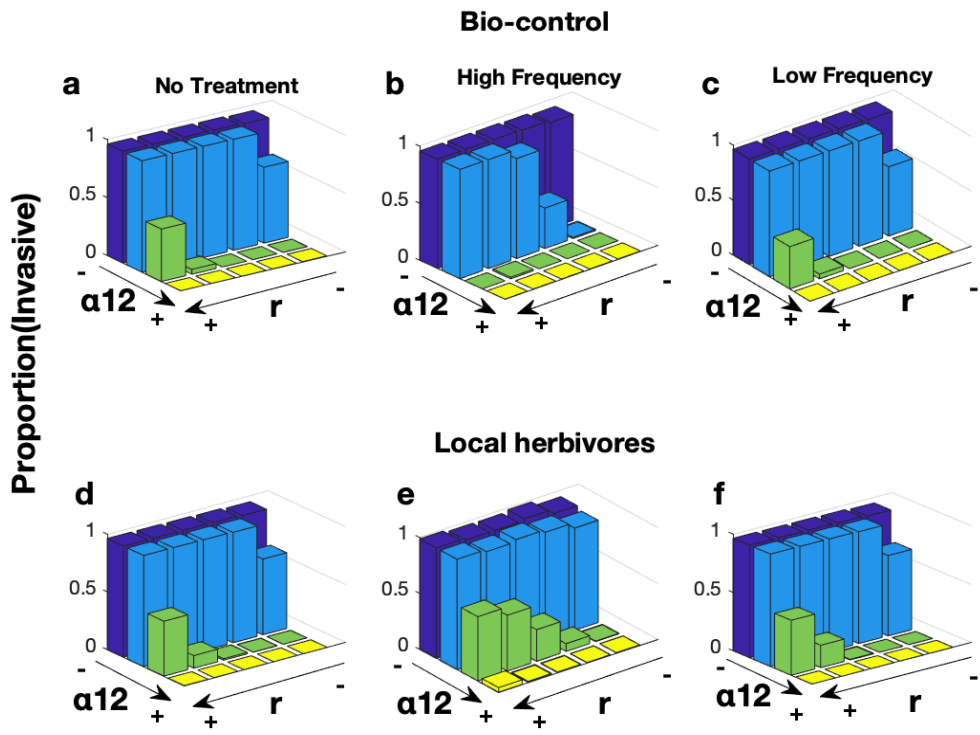

Figure 5

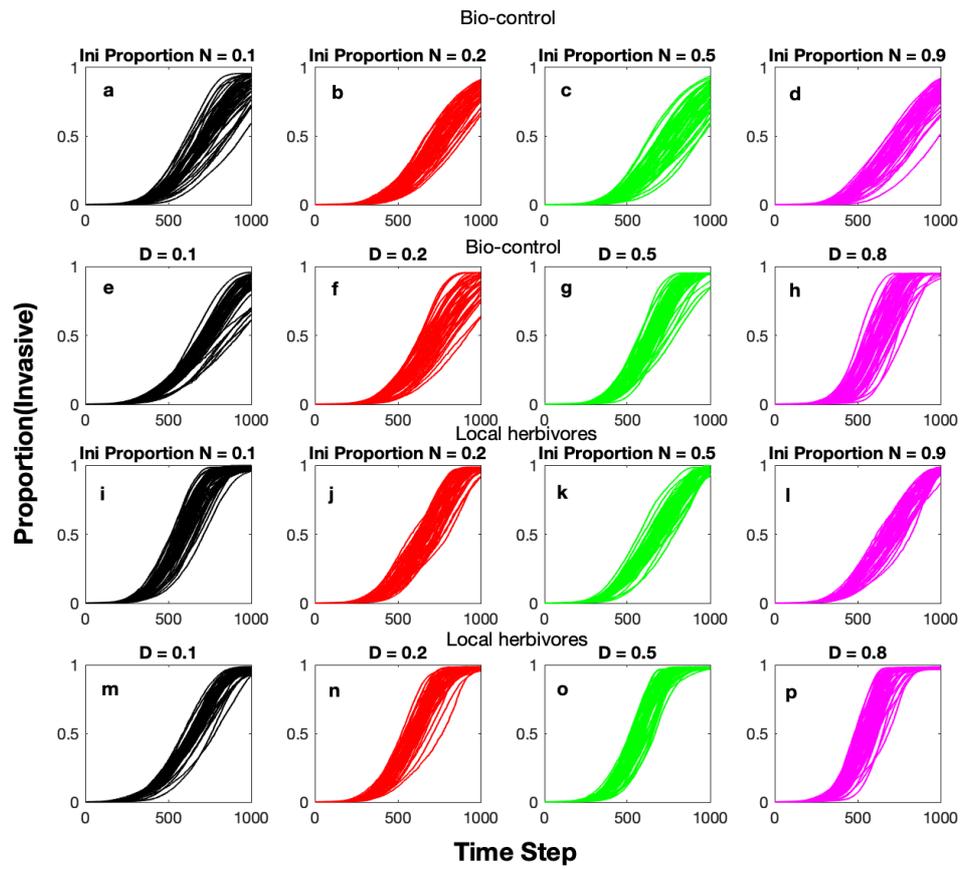

Table 1 Parameters used in the model

| Description | Native | | Invasive | |
|---|---|---|---|---|
| | Symbol | Default value | Symbol | Default value |
| Intrinsic growth rate | $r$ | *0.2* | $R$ | *0.2* |
| Competition coefficient | $\alpha_{12}$ | *1* | $\alpha_{21}$ | *2* |
| Dispersal rate | $d$ | *0.1* | $D$ | *0.2* |
| Initial population | $N_0$ | *Random (0-1)* | $I_0$ | *2* |

| | Plot | |
|---|---|---|
| | Symbol | Default value |
| Carrying capacity | $K$ | *10/cell* |
| Row | $i$ | *100* |
| Column | $j$ | *100* |